\documentclass[preprintnumbers,aps,prd,twocolumn,showpacs,showkeys,floatfix,preprintnumbers,letterpaper,amsmath,amssymb,superscriptaddress]{revtex4-1}
\usepackage{subfigure}
\usepackage{graphicx}
\usepackage{dcolumn}
\usepackage{epsfig}
\usepackage{amsmath, bm}
\usepackage{amsfonts}
\usepackage{amssymb}
\usepackage{color} 
\usepackage{xcolor}
\usepackage{hyperref}
\usepackage{tabularx}
\usepackage{float}
\usepackage[normalem]{ulem}

\newcommand{\mnras}{Mon. Not. R. Astron. Soc.}

\newcommand{\jcap}{J. Cosmol. Astropart. Phys.}

\usepackage[T1]{fontenc}
\usepackage{ae,aecompl}
\newcommand{\be}{\begin{equation}}
\newcommand{\ee}{\end{equation}}
\newcommand{\bea}{\begin{eqnarray}}
\newcommand{\eea}{\end{eqnarray}}

\begin{document}

\title{Displacement Field Analysis via Optimal Transport:\\ Multi-Tracer Approach to Cosmological Reconstruction}

\author{Farnik Nikakhtar}
\email{farnik.nikakhtar@yale.edu}
\affiliation{Department of Physics, Yale University, New Haven, CT 06511, USA}

\author{Ravi K.~Sheth}
\affiliation{Center for Particle Cosmology, University of Pennsylvania, Philadelphia, PA 19104, USA}
\affiliation{The Abdus Salam International Center for Theoretical Physics, Strada Costiera 11, Trieste 34151, Italy}

\author{Nikhil Padmanabhan}
\affiliation{Department of Physics, Yale University, New Haven, CT 06511, USA} 
\affiliation{Department of Astronomy, Yale University, New Haven, CT 06511, USA}

\author{Bruno L\'evy}
\affiliation{Centre Inria de Saclay, Universit\'e Paris Saclay, Laboratoire de Math\'ematiques d'Orsay}

\author{Roya Mohayaee}
\affiliation{Sorbonne Universit\'e, CNRS, Institut d'Astrophysique de Paris, 98bis Bld Arago, 75014 Paris, France}
\affiliation{Rudolf Peierls Centre for Theoretical Physics, University of Oxford, Parks Road, Oxford OX1 3PU, United Kingdom}

\date{\today}

\begin{abstract}
We demonstrate the effectiveness of one of the many multi-tracer analyses enabled by Optimal Transport (OT) reconstruction. Leveraging a semi-discrete OT algorithm, we determine the displacements between initial and observed positions of biased tracers and the remaining matter field. With only redshift-space distorted final positions of biased tracers and a simple premise for the remaining mass distribution as input, OT solves the displacement field. This extracted field, assuming asymptotically uniform density and a gradient flow displacement, enables reconstruction of the initial overdensity fluctuation field. We show that the divergence of the OT displacement field is a good proxy of the linear density field, even though the method never assumes the linear theory growth. Additionally, this divergence field can be combined with the reconstructed protohalos to provide a higher signal-to-noise measurement of the BAO standard ruler than was possible with either measurement individually.
\end{abstract}


\maketitle


\section{Introduction}\label{intro}

Essentially all currently viable models for cosmological structure formation suggest that the initial conditions were almost exquisitely uniform.  In fact, in simulations of cosmological structure formation, the initial density field is homogeneous (particles start from a grid), and subsequent evolution is driven by inhomogeneous perturbations in the initial velocity field.  In this case, the initial velocity field is the gradient of a potential field, and the overdensity field is explicitly related to the divergence of the initial velocity field:  
\begin{equation}
    \delta_{init}\propto \nabla\cdot \bm{v}_{init} \propto \nabla^2\Phi_{init},
    \label{eq:div}
\end{equation}
where $\Phi_{init}$ is the initial gravitational potential \cite{Peebles:1980}. Although the constant of proportionality depends on the cosmological model, the relation is otherwise general.  

It is common to use $\bm{q}$ to denote the initial position of a particle, and $\bm{x}$ its position at some later time $t$.  Then the {\em displacement} $\bm{S}$ is defined by 
\begin{equation}
 \bm{x}(t) = \bm{q} + \bm{S}(\bm{q},t).
 \label{eq:xqSq}
\end{equation} 
Optimal Transport (hereafter OT) theory is a fast growing gem in mathematics that, broadly, computes distances and relations between different probability distributions \cite{villani}. In cosmology OT is a natural framework to use since it can \emph{exactly} reconstruct the displacements $\bm{S}(\bm{q},t)$ in \eqref{eq:xqSq} if they are gradients of a potential (see \cite{EUR, BrenierPFMR91,BenamouBrenier} and references therein).  In cosmological structure formation, this is explicitly true in Lagrangian Perturbation theory upto and including the second order (see {e.g.} \cite{buchert1994}).  This, and the fact that the initial conditions were `uniform' has motivated the use of OT methods for reconstructing the displacements, from knowledge of only the particle positions at some late time \cite{2002Natur.417..260F,EUR}.  {\it i.e.} if $n$ labels a particle, then OT determines estimates $\bm{S}_{{\rm OT}n}$ of each of the displacements $\bm{S}_n$, knowing only the set of final positions $\bm{x}_n$.  From this, $\bm{q}_{{\rm OT}n} \equiv \bm{x}_n - \bm{S}_{{\rm OT}n}$.  Recent development of semi-discrete optimal transport methods \cite{journals/M2AN/LevyNAL15, KMT2019} has transformed the field by vastly increasing the speed with which the displacements, and hence the initial conditions, can be reconstructed \cite{PRLdm, royaMAK, PRLhalos}. Note, the displacement $\bm{S}$ can be equivalently thought of as a function of $\bm{q}$ or $\bm{x}$. In computing derivatives, spatial derivatives with respect to $\bm{q}$ will be denoted as $\nabla_{\bm{q}}$, while those with respect to $\bm{x}$ will be denoted as $\nabla_{\bm{x}}$.

Such reconstructions are particularly attractive for studies which seek to constrain cosmological parameters from the statistical properties of the initial field \cite{recSDSS}.  The most used summary statistic of the field is the power spectrum $P(k)$, or its Fourier transform, the two-point correlation function $\xi(r)$.  Typically, these are estimated directly from the particle distribution.  However, the distribution of $\bm{q}_{{\rm OT}n}$ is uniform (by design!), so $\xi_{\rm OT}(r)=0$ for all $r$.  For this reason, Ref.\cite{PRLdm} estimated two-point statistics of the slightly evolved positions $\bm{x}_n(\alpha) = \bm{q}_{{\rm OT}n} + \alpha\,\bm{S}_{{\rm OT}n}$, for $\alpha\sim 0.1$.  To compare this with the `initial' conditions in the simulations, they used the evolved positions $\bm{q}_n + \bm{v}_{init-n}/(afH)$ (the displacement is just initial speed times a suitably chosen time), but because $afH = a\, (d\ln D/d\ln a)(d\ln a/dt)$ depends on a cosmological model ($D(a)$ is the linear theory growth factor), matching $\alpha$ to $(afH)^{-1}$ gave the impression that the comparison was model dependent.

One of the goals of this work is to show that equation~(\ref{eq:div}) offers a more direct, cosmology independent estimate of the statistics of the initial fluctuation field.  (Strictly speaking, to a good approximation, it is the initial fluctuation field evolved using linear theory to the time of observation.)  We do this in two steps.  In Section~\ref{sec:ideal}, we consider the ideal case in which one starts from perfect information about the evolved dark matter field (all the $\bm{x}_n$ are known), as in Ref.\cite{PRLdm}.  In Section~\ref{sec:wiener}, we show how to treat the case in which evolved positions of only a biased subset of the full field are known, as in Ref.\cite{PRLhalos}.  This second step requires a guess about the distribution of the mass that is not directly associated with the biased tracers, which we refer to as the `dust'.  

While previous work in the OT context primarily focused on reconstructing the biased field \cite{OTrsd}, here we show the dual reconstruction capability inherent in Optimal Transport (OT) - namely, the reconstruction of both the protohalo field and the linear dark matter field. This leads to a discussion on the potential synergy between these reconstructions, providing a good context for contrasting our reconstruction scheme with other approaches. In addition, Section~\ref{sec:multi} discusses our results in the context of multi-tracer studies of large-scale structure, and highlights the fact that OT-like methods allow displacements and quantities built from the displacement field to be used as tracers of the cosmological fluctuation field.  We summarize our results in Section~\ref{sec:discussion}. 

We demonstrate our methods using the dark matter and halo distributions in the same HADES cosmological simulations \cite{hades} that were used in Ref.\cite{OTrsd}.  The background cosmological model is flat $\Lambda$CDM with $(\Omega_m,\Omega_b,h) = 0.3175, 0.049, 0.6711)$.  The initial fluctuations were Gaussian with $P_{\rm Lin}(k)$ having shape and amplitude parameters $(n_s,\sigma_8)=0.9624,0.833)$, and each simulation evolves $512^3$ particles in a periodic box that is $L=1h^{-1}$Gpc on a side, so the particle mass is $m_p = 6.5\times 10^{11}h^{-1}M_\odot$.  We focus on the $z=0$ outputs, where we consider the dark matter field, as well as halos more massive than $20m_p$;  see Ref.\cite{OTrsd} for further details about how the halos were identified.  All our results are averaged over 20 simulation boxes.  

\section{Perfect knowledge of the full nonlinear field}\label{sec:ideal}

\subsection{Divergence of displacement}
Before we begin, it is important to distinguish between a number of measures of the initial power spectrum (PS).  First is the theoretical PS which is used to generate the Fourier-modes of the initial fluctuation field $\delta_{init}$.  We refer to the related power spectrum, scaled to $z=0$ using linear theory, as $P_{\rm Lin}(k)$. We use $P_{\rm IC}(k)$ to denote the power spectrum of the scalar $\delta_{\rm IC}$, which is the initial fluctuation field $\delta_{init}$, scaled to $z=0$ using linear theory. This power spectrum should differ from $P_{\rm Lin}(k)$ only because it is a single stochastic realization (of which $P_{\rm Lin}$ is the mean).

The Fourier modes of the initial velocities and densities are related: 
 ${\bm v}_{init} = {\rm i}{\bm k}\, \delta_{init}/k^2$,
so that 
$\nabla\cdot {\bm v}_{init} = - (afH)_{init}\,\delta_{init}$. The vector ${\bm v}_{init}/(afH)_{init}$ has units of distance, and $(D_0/D_{init})\,{\bm v}_{init}/(afH)_{init}\equiv {\bm S}_{\rm Zel}$ corresponds to the displacement vector, from the initial to the present time, in the Zeldovich approximation.  We will define the scalar 
 $\Theta_{\rm Zel}\equiv \nabla_{\bm q}\cdot {\bm S}_{\rm Zel}$, 
and use $P_{\Theta_{\rm Z}\Theta_{\rm Z}}(k)$ to denote the power spectrum of $\Theta_{\rm Zel}$.  
Since $\Theta_{\rm Zel} = - (D_0/D_{init})\,\delta_{init} = - \delta_{\rm IC}$, $P_{\Theta_{\rm Z}\Theta_{\rm Z}}(k)$ should be the same as $P_{\rm IC}(k)$, so a comparison of the two measures the accuracy of our scheme for estimating velocity divergences in the simulations.  

Note that $P_{\Theta_{\rm Z}\Theta_{\rm Z}}(k)$ will differ from the power spectrum of the Zeldovich-evolved density field, where particles move along straight lines following the gradient flow until $z=0$, as detailed in \cite{zelPk}.  In addition, since the true displacement of a particle, ${\bm S}_{true}$, is different from the approximate ${\bm S}_{\rm Zel}$ (the Zeldovich displacement is just the first order term in Lagrangian Perturbation theory), $P_{\Theta_{\rm Z}\Theta_{\rm Z}}$ differs from the power spectrum of the divergence of the actual displacements, $P_{\Theta\Theta}(k)$, where 
\be
\Theta\equiv\nabla_{\bm q} \cdot {\bm S}_{true}\,
\ee
and we shall discuss $\nabla_{\bm x} \cdot {\bm S}$, the divergence at $\bm{x}$ rather than $\bm{q}$, in Appendix C.

Finally, we define $\Theta_{\rm OT}$ as the divergence of the displacements returned by our OT reconstructions.  Recall that whereas the set of $\bm{S}_{true} = \bm{x} - \bm{q}$ displacements uses a full cosmological simulation to obtain the evolved positions $\bm{x}$ from an initial grid of $\bm{q}$ positions, OT starts from this list of $\bm{x}$ positions, and returns a list of corresponding $\bm{q}$s (Laguerre cell barycenters that are not necessarily on a grid).  

\begin{figure}
    \centering
    \includegraphics[width=\linewidth]{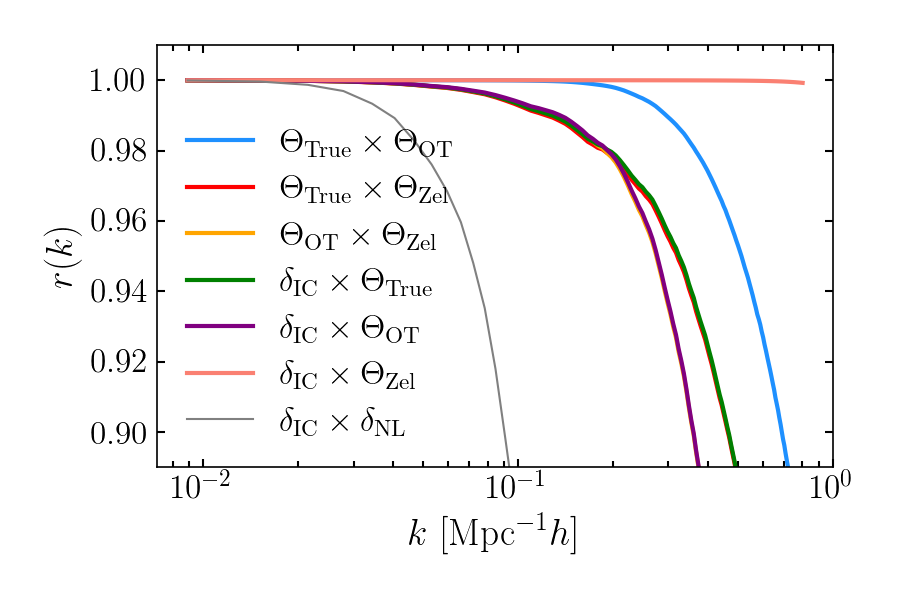}
    \caption{Normalized correlation coefficients between the various divergences in real space, as defined in the text.  On the $k\le 0.2h$/Mpc scales that are most relevant to BAO analyses, the OT displacement divergences are remarkably well correlated with the true ones (blue), even though the OT algorithm includes no cosmological information. They are less well correlated with the initial density field (purple that overlaps with yellow), but are still much better than the nonlinear density (grey). The green curve overlaps with the red curve as well.  
    \label{fig:rk}}
\end{figure}

\subsection{Cross-correlations}
Before we look at the power spectra themselves, it is helpful to look at how closely these different quantities correlate with one another.  To quantify the degree of correlation, we use the normalized coefficient 
\begin{equation}
    r_{AB}(k) \equiv \frac{\langle P_{AB}(k)\rangle}{\sqrt{P_{AA}(k)P_{BB}(k)}}.
\end{equation}
(Note that we do not subtract shotnoise from any measurements.) To help calibrate the figures throughout the paper, the thin grey curve in Figure~\ref{fig:rk} shows that the nonlinear field is well-correlated with the initial one only on very large scales ($r\ge 0.9$ only if $k\le 0.1h$Mpc$^{-1}$).  We see that $\delta_{init}$ and $\Theta_{\rm Z}$ are extremely tightly correlated (pink curve shows $r\sim 1$ for essentially all $k$), as expected.  More interesting is that $\Theta$ and $\Theta_{\rm Z}$, and $\Theta$ and $\delta_{init}$, are also quite well correlated on BAO scales (red and green curves, which overlap almost exactly, have $r\ge 0.9$ for $k\le 0.45\,h$Mpc$^{-1}$).  This matters, because the reconstructed $\Theta_{\rm OT}$ is extremely well correlated with the true $\Theta$ (blue curve) -- this, after all, is why OT-reconstruction is interesting in the first place -- so $\Theta_{\rm OT}$ is well correlated with both $\Theta_{\rm Z}$ and $\delta_{init}$ (yellow and purple curves, which overlap almost exactly, show $r\ge 0.9$ for $k\le 0.35\,h$Mpc$^{-1}$).  Hence, compared to the nonlinear density field, the reconstructed $\Theta_{\rm OT}$ field provides about $(0.35/0.1)^3 \sim 40\times$ more $k$-modes with $r \sim 1$ to the initial fluctuation field.  

This is non-trivial, because the explicit goal of OT is to return $\bm{S}_{\rm OT} = \bm{S}_{true}$, and the blue curve in Figure~\ref{fig:rk} suggests that it does this rather well.  However, other than doing so while also returning a uniform density field, it does not explicitly try to reconstruct $\delta_{init}$ or $\Theta_{\rm Z}$.  Since $\bm{S}_{true}\ne \bm{S}_{\rm Zel}$, even if OT returned $\bm{S}_{true}$ perfectly, it would not perform better -- in the sense of correlating with $\delta_{init}$ -- than the green curve shown in Figure~\ref{fig:rk}.

Stated differently:  if we use $\Phi_{\rm OT}(\bm{q})$ to denote the potential which determines the OT reconstructed displacements $\bm{\nabla} \Phi_{\rm OT}$, then, because $\bm{S}_{\rm OT}\approx\bm{S}_{true}\ne \bm{S}_{\rm Zel}$, we know that $\Phi_{\rm OT}\ne\Phi_{init}$. This is not surprising, since by definition, the gradient of the OT potential moves particles to their exact nonlinear positions, in contrast to $\bm{\nabla} \Phi_{init}$, which would result in the Zeldovich-evolved field. Nevertheless, the correlation between $\Theta_{\rm OT}\equiv k^2\Phi_{\rm OT}$ and $\delta_{init}\equiv k^2\Phi_{init}$ (purple curve) is sufficiently good on BAO scales that we should be able to use $\Theta_{\rm OT}$ as a proxy for $\delta_{init}$.  We test this explicitly in the next sections.

\begin{figure}
    \centering
    \includegraphics[width=\linewidth]{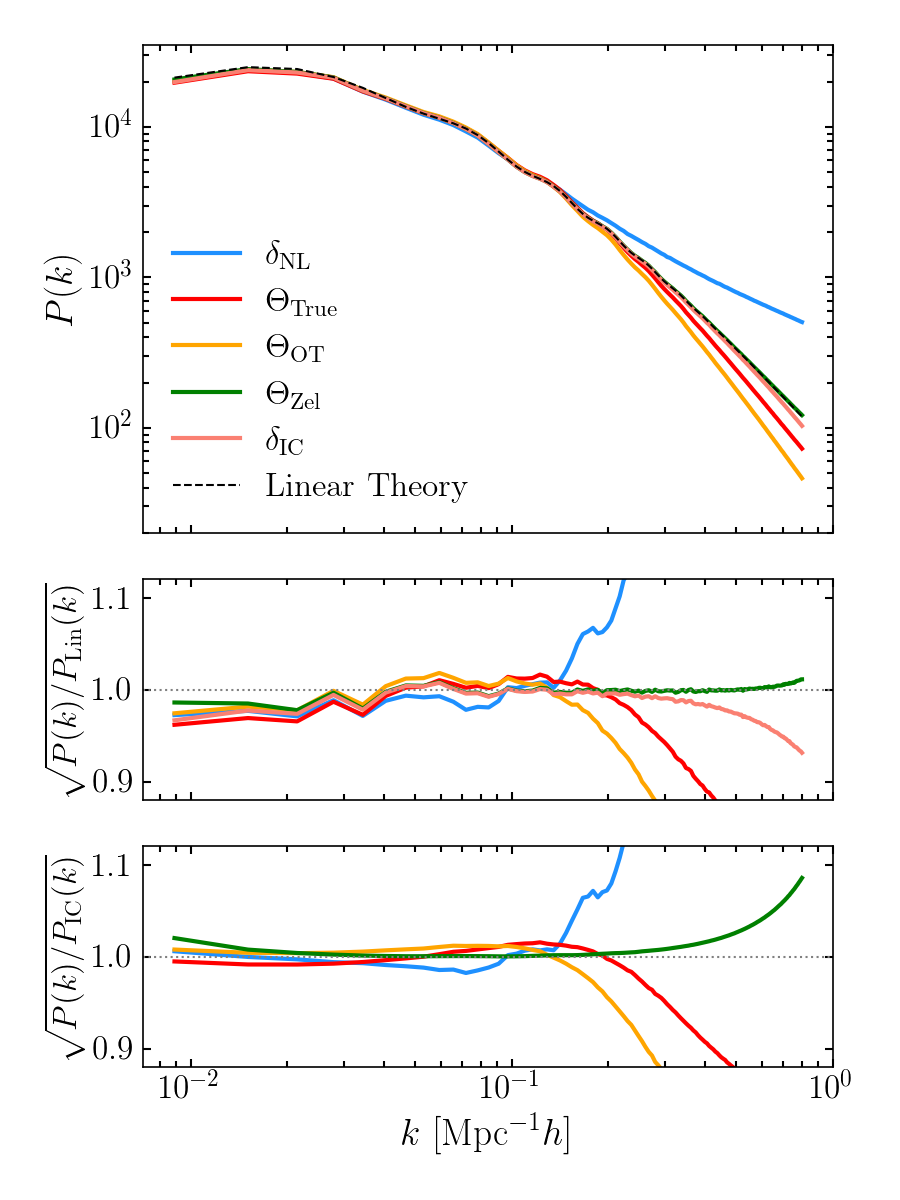}
    \caption{Comparison of various power spectra in real space (as labeled).  Dashed line shows the expected $z=0$ linear theory curve, $P_{\rm Lin}(k)$, and upper-most blue line shows the actual, evolved, nonlinear curve $P_{\rm NL}(k)$.  Yellow curve, for $\Theta_{\rm OT}$, represents our reconstruction of the shape and amplitude of $P_{\rm Lin}(k)$ from $P_{\rm NL}(k)$, despite having no cosmological information (e.g. linear theory growth factor).  Middle panel shows the result of dividing each of these curves by $P_{\rm Lin}(k)$ and taking the square root.  In bottom panel, they were each divided by $P_{\rm IC}(k)$; this removes cosmic variance. 
    \label{fig:Pk}}
\end{figure}

\subsection{Power-spectra and correlation functions}
Figure~\ref{fig:Pk} shows the power spectra associated with these different divergences.  To guide the eye, the dashed curve in the top panel shows $P_{\rm Lin}(k)$; it is very similar to $P_{\rm IC}$ and to $P_{\Theta_{\rm Z}\Theta_{\rm Z}}(k)$.  The power spectrum of true divergences is also remarkably similar to $P_{\rm Lin}(k)$, despite the fact that $\Theta$ is not as well correlated with $\delta_{init}$.  

We are most interested in $\Theta_{\rm OT}$.  Although the curve for $\Theta_{\rm OT}$ has noticeably less power than the others at large $k$, it is in excellent agreement on BAO scales.  Note in particular that, whereas amplitudes are scaled out in $r(k)$, they matter here.  Nevertheless, on BAO scales, the power spectrum of $\Theta_{\rm OT}$ is an excellent tracer of $P_{\rm Lin}(k)$.  To emphasize the agreement, the middle panel shows the (square root of the) ratio of these $P(k)$ to $P_{\rm Lin}(k)$.  This shows that $\Theta_{\rm OT}$ is within a few percent of linear theory on BAO scales; in fact, it traces the jaggedness at small $k$ that is due to cosmic variance.  The bottom panel shows the result of taking the ratio with respect to $P_{\Theta_{\rm Z}\Theta_{\rm Z}}(k)$ instead (we take the average of the ratio, rather than the ratio of the averages).  This removes the cosmic variance, and shows that $\Theta_{\rm OT}(k)$ has the linear theory amplitude to within a few percent for $k\le 0.1h$Mpc$^{-1}$.  (While the nonlinear field also has the right amplitude at low $k$, the previous figure shows that it has $r(k)\sim 1$ over a much smaller range of $k$.)

\begin{figure}
    \centering
    \includegraphics[width=\linewidth]{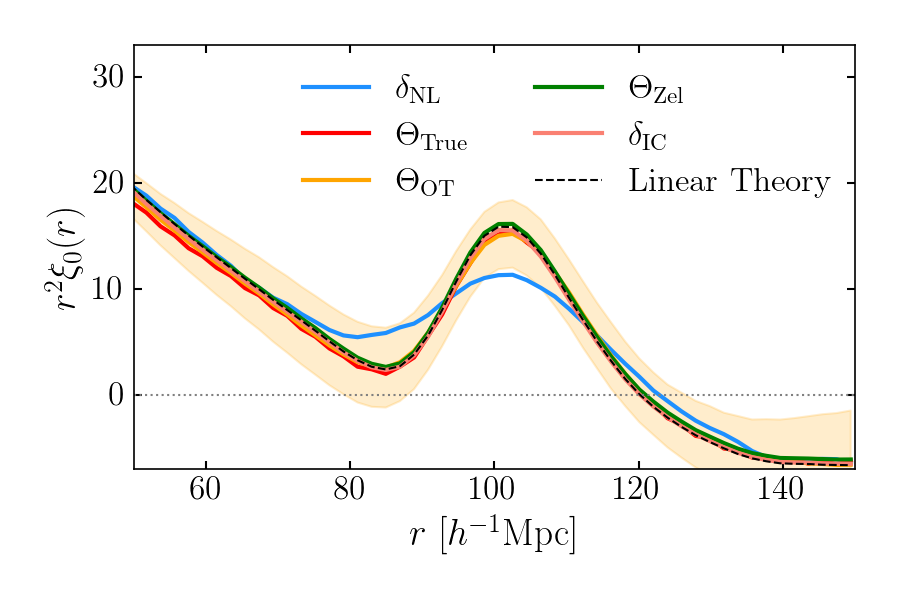}
    \caption{Correlation functions corresponding to the power spectra shown in the previous figure.  Note in particular that it was not necessary to specify a growth factor for the $\Theta_{\rm OT}$ curve, but it is, nevertheless, indistinguishable from the other linear field curves. 
    \label{fig:xir}}
\end{figure}

Figure~\ref{fig:xir} shows the result of Fourier transforming the power spectra shown in the previous figure. In practice, this transformation is performed using a Fast Fourier Transform (FFT) on the 3D power spectrum calculated in the box. On BAO scales, all the curves are in excellent agreement with linear theory (shaded region shows the rms scatter between 20 simulations), except for the one labeled nonlinear, for which the BAO feature is obviously smeared-out.  In particular, $\xi$ of $\Theta_{\rm OT}$ reproduces the shape and amplitude of $\xi_{\rm Lin}$ (on BAO scales) rather well.  This demonstrates how one can use the OT displacements to reconstruct the shape and amplitude of $\xi_{\rm Lin}$ (on BAO scales) without using any cosmological information.  In particular, it was not necessary to specify a growth factor when defining the reconstructed displacements, and the extra scaling used by Ref.\cite{PRLdm} is unnecessary.

Before we end this section, we recall from the Introduction that one can think of OT as {\em approximately} returning the initial fluctuation field evolved using linear theory to the time of observation.  This is because, although our results show that this approximation is very good, the OT reconstruction does differ slightly from $\xi_{\rm Lin}$. We know that $\Phi_{\rm OT}$ is not the linear $\Phi$, but has higher order corrections to it. It may be possible to construct a better estimate of the linear $\Phi$ from the $\Phi_{\rm OT}$. Exploring this is currently work in progress.

\section{Biased, redshift-space distorted tracers}\label{sec:wiener}
In the previous section we have outlined what could be achieved if one had a perfect knowledge of positions (but not velocities) in the nonlinear field.  In practice, one usually only has information about a biased subset of the nonlinear field whose positions have been distorted by peculiar velocities, so we now discuss how to extend our method to this more realistic case.  

\subsection{Reconstructing both biased tracers and the total field}
The OT approach assumes that a reasonably accurate mass estimate is available for each tracer, and all statistics are `mass-weighted'.  However, it also requires knowledge of the amount and spatial distribution of the mass that is not observed, what we call the `dust'.  Ref.~\cite{OTrsd} shows that in order to model the missing dust one must assume values for two cosmological parameters:  $\Omega_m$ and $\sigma_8$.  

\begin{figure}
    \centering
    \includegraphics[width=\linewidth]{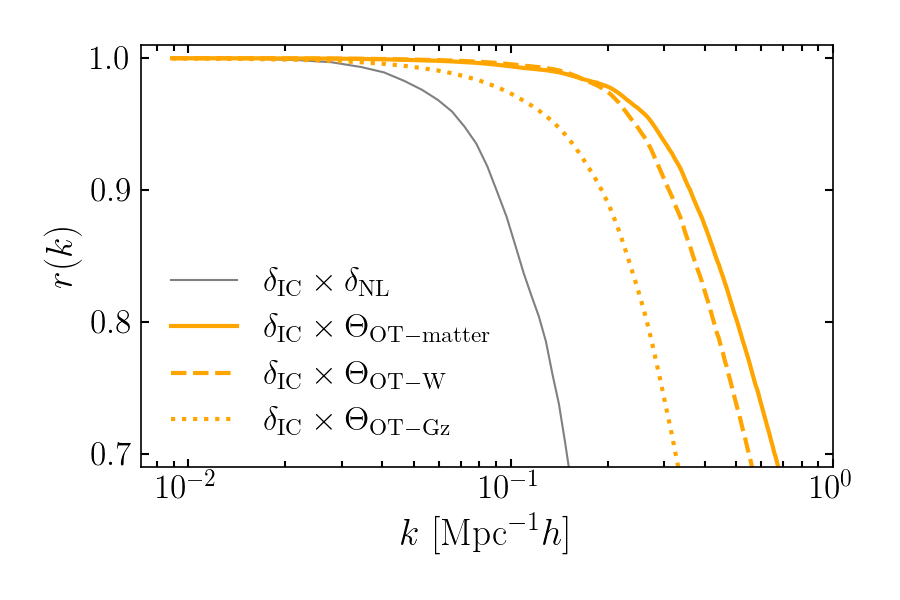}
    \caption{Dependence of normalized cross-correlation between the full initial and reconstructed fields, when the OT algorithm is started from the full nonlinear field (solid; OT-matter), from mass-weighted halos plus the true dust in real-space (dashed; OT-W), and from mass-weighted halos plus the Wiener filter model for the dust, in redshift-space (dotted; OT-Gz).  For comparison, the grey curve (same as in Figure~\ref{fig:rk}) shows the  $\delta_{\rm IC}$-$\delta_{\rm NL}$ correlation.
    \label{fig:rkWGz}}
\end{figure}

The curve that is second from top in Figure~\ref{fig:rkWGz} shows the cross-correlation between the full initial and reconstructed fields when the OT algorithm is started with mass-weighted halos more massive than $10^{13}h^{-1}M_\odot$ plus the true dust (i.e. particles that were not in these halos) in real-space (dashed).  For comparison, the top (solid) curve shows the result of starting from the full nonlinear field (i.e. same as the yellow or purple curves in Figure~\ref{fig:rk}). Some of the degradation in $r(k)$ is because, in our reconstruction, we treat all the mass of a halo as being concentrated at its centre, and the mass of the reconstructed protohalo as being located at the barycenter of its Laguerre cell, so that all of a halo's mass has the same displacement vector:  $\bm{x}_{\rm com} - \bm{q}_{\rm com}$.  A better model would distribute the mass uniformly within the Laguerre cell, with displacement given by $(\bm{x}_{\rm com} - \bm{q}_{\rm com}) - (\bm{q}-\bm{q}_{\rm com})$.  

Of course, in practice, we do not know the true spatial distribution of the dust.  If our guess is not accurate, then this will further degrade $r(k)$.  Ref.~\cite{OTrsd} discusses a number of approximations which one can use to model the dust.  A simple Wiener filter model, that uses the observed tracers to guide the placement of the dust particles, was shown to work quite well and shall hence be used hereafter.  For this model, if $\delta_b$ is the fluctuation in the biased tracer field, then 
\begin{equation}
 \delta_d = \frac{\langle\delta_b\delta_d\rangle}{\langle\delta_b\delta_b\rangle}\,\delta_b
  \approx (b_d/b)\,\delta_b,
 \label{eq:wienerdust}
\end{equation}
where the final expression assumes that 
 $\langle\delta_b\delta_d\rangle \approx bb_d\,P(k)$ 
and 
 $\langle\delta_b^2\rangle \approx b^2\,P(k) + {\rm noise} \approx b^2\,P(k)$ 
on BAO scales.

In addition, Ref.~\cite{OTrsd} also studied the impact of redshift-space distortions -- the anisotropy induced by the fact that estimated distances along the line of sight are distorted by peculiar velocities -- finding that ignoring them altogether was a rather good approximation, at least for restoring the monopole of the two-point correlation to its initial shape.  

\begin{figure}
    \centering
    \includegraphics[width=\linewidth]{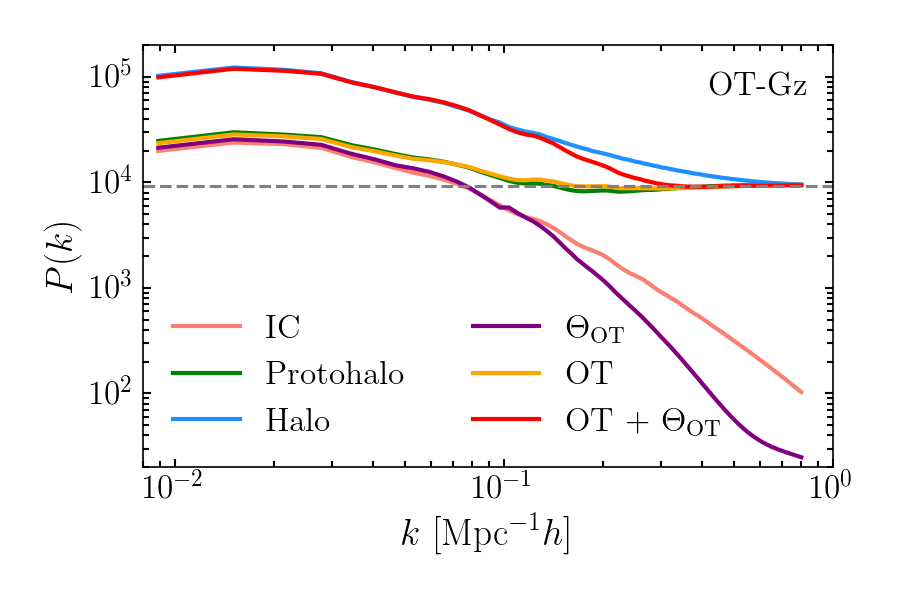}
    \caption{Initial, evolved and reconstructed power spectra.  Top blue curve shows observed monopole of a biased, redshift-space distorted set of tracers (mass-weighted halos more massive than $10^{13}h^{-1}M_\odot$); green and peach curves show the power spectra of the corresponding protohalos and of the initial dark matter field ($P_{\rm IC}\approx P_{\rm Lin}$).  Yellow and purple curves show the power spectra of the OT reconstructed halos and $\Theta_{\rm OT}$ of the full reconstructed field, and red curve shows the power spectrum of the field which is their sum (OT reconstructed halo fluctuations plus $\Theta_{\rm OT}$). Dashed black line shows the shot-noise for mass-weighted halos.
    \label{fig:PkGz}}
\end{figure}

The curve that is third from top in Figure~\ref{fig:rkWGz} (dotted) shows the result of starting from redshift-space distorted positions of biased tracers, using the Wiener filter model to generate the dust, performing the semi-discrete OT reconstruction, and then measuring $r(k)$ between $\Theta_{\rm OT}$ of the full field (i.e. reconstructed halos plus dust) and $\delta_{init}$.  Evidently, inaccuracies in the Wiener-filter model for the dust and redshift space distortions do degrade $r(k)$, but it is still significantly better than $\delta_{\rm NL}$ (bottom grey curve, same as the grey curve in Figure~\ref{fig:rk}).  

\begin{figure}
    \centering
    \includegraphics[width=\linewidth]{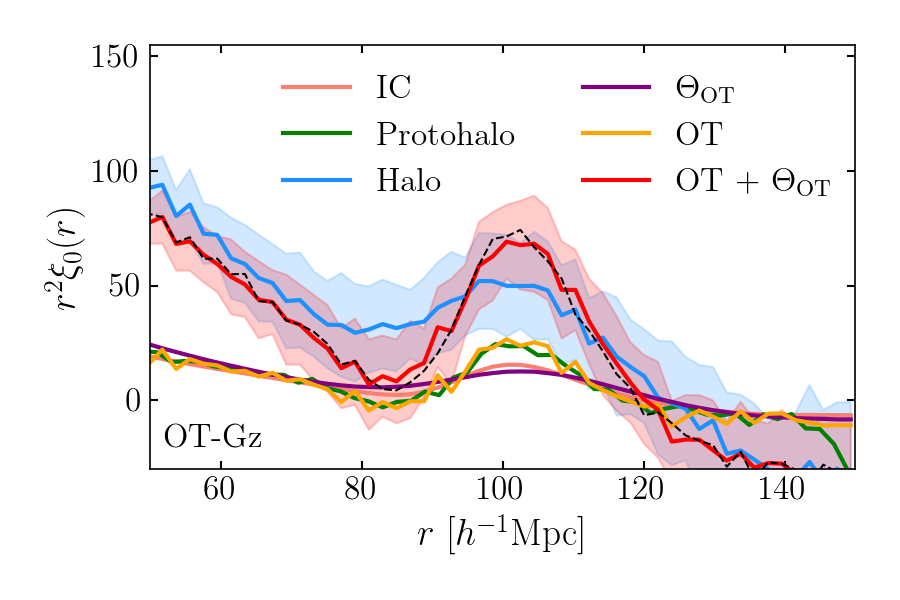}
    \caption{Correlation functions corresponding to the power spectra shown in Figure~\ref{fig:PkGz}. The dashed black curve shows the correlation function of the sum of the actual protohalo and $\delta_{init}$ fluctuation fields.
    \label{fig:xiGz}}
\end{figure}

Figures~\ref{fig:PkGz} and~\ref{fig:xiGz} show the corresponding $P(k)$ and $\xi(r)$ of the reconstructed protohalos and of $\Theta_{\rm OT}$. We do not subtract shot-noise from the measured $P(k)$; this is negligible for the full field, but significant for the halos and protohalos.  
(We mass-weight the halos, so the shot-noise is not $\bar{n}_h^{-1}$, but 
$(\sum_i m_i^2/V)/ (\sum_i m_i/V)^2 
  = \bar{n}_h^{-1} \langle m^2\rangle / \langle m\rangle^2$.)
 
The first thing to notice is that $P(k)$ and $\xi(r)$ of the OT reconstructed halos (yellow) are extremely similar to the protohalos (green).  This is the agreement previously reported by Ref.~\cite{OTrsd}.  Although Ref.~\cite{OTrsd} focused on the reconstructed position of biased tracers, the same OT algorithm also outputs the displacements of the dust particles as well. We can use the full set of displacements (mass-weighted halos plus dust) to estimate $\Theta_{\rm OT}$ as described in the previous section.  This gives the purple curves, which are very close to $P_{\rm IC}$ or $\xi_{\rm IC}$ (pink) on BAO scales.  Evidently, neither inaccuracies in the Wiener-filter model for the dust nor redshift space distortions degrade the agreement substantially compared to the ideal case shown in the previous section.  We conclude that our OT methodology successfully reconstructs both the positions of the biased tracers {\em and} the fluctuations in the full field.  

\subsection{Combining the two reconstructions}\label{sec:combine} 
Figures~\ref{fig:PkGz} and~\ref{fig:xiGz} show that the amplitude of the clustering signal of the reconstructed biased tracers is smaller than that in the evolved field.  The physics which drives this difference -- essentially the continuity equation -- dictates that if the biased subset initially satisfies $\delta_b = b_{\rm L}\delta_{\rm L}$ then the evolved subset satisfies $(b_{\rm L}+1)\delta_{\rm NL}$ (e.g., Ref.\cite{mw1996, PRLhalos}). However, because the amplitude of the reconstructed signal is smaller (yellow curve lies far below blue), the signal to noise of the BAO feature in the reconstructed field is diminished.  We can use our knowledge of the physics to correct for this as follows.  

On BAO scales, $\delta_{\rm NL}\approx \delta_{\rm L}\approx \Theta_{\rm Zel}$, so the evolved subset $(b_{\rm L}+1)\delta_{\rm NL}$ is approximately $b_{\rm L}\delta_{\rm L} + \Theta_{\rm Zel}$.  This suggests that if we add the reconstructed halo and $\Theta_{\rm OT}$ fluctuation fields, then the result should have a power spectrum which is given by
\begin{equation}
\langle (b_{\rm L}\delta_{\rm L} + \Theta_{\rm Zel})^2\rangle 
 \approx b_{\rm L}^2P_{\rm Lin} + 2b_{\rm L}P_{\rm Lin} + P_{\rm Lin}
 \approx (b + 1)^2\,P_{\rm Lin}. 
\end{equation}
I.e., it should have approximately the same amplitude as the original signal.  The red curve shows this result:  The overall amplitude of the signal is nearly restored.  On scales slightly below the BAO feature, the red curve lies slightly below the blue one.  This slight mismatch is reassuring, because the blue curve is the redshift-space distorted monopole, so it is slightly enhanced compared to the real-space correlation \cite{kaiser1987}, which is the quantity that OT aims to reconstruct.  

In addition to this slight difference in amplitude, the OT reconstructed signal also has a sharper BAO feature compared to the linear matter field. The reason for this is also understood \cite{PRLhalos}:  most protohalos are  associated with peaks in the initial field, and this is due to the scale-dependent nature of the `bias' factor $b_{\rm L}$, which includes a $k^2$ term that sharpens the peak amplitude \cite{bbks86, rsdPeaks}.  There is further corroborating evidence of the peak-protohalo correspondence: the initial speeds of peaks are predicted to be smaller than the rms speeds of randomly chosen positions, and protohalo speeds are indeed smaller, and well reproduced by peaks theory \cite{sd2001}.  So it is reassuring that OT returns smaller rms displacements for the halo field ($\sim 9.1h^{-1}$Mpc) than for the total ($9.8h^{-1}$Mpc).  

In the evolved field, the scale independent piece increases, but the scale dependent piece decreases (both by unity) \cite{bkPeaks}.  Adding the linear (rather than the evolved) fluctuation field to that of the protohalos increases the scale independent bias while leaving the scale dependent part unchanged.  This results in an enhanced BAO signal.  To check that this is indeed what is going on, the dashed black curve shows the correlation function of the field that is the sum of the actual protohalo and $\delta_{init}$ fluctuation fields.  Our (red) OT-reconstructed curve is in great agreement, indicating that the reason for the enhanced signal is understood.  

Of course, the two reconstructed fields are not independent, so when constraining parameters from the shape/scale of the BAO signal, the covariance between the reconstructed halo and matter (i.e. $\Theta_{\rm OT}$) fields must be taken into account.  This raises the question:  What combination (linear or nonlinear) of the two fields has the most constraining power ?  In this context, the straight linear sum of the two fields is not obviously optimal, but it does illustrate the potential gains from combining the two.  We return to this in the next section.

\subsection{Relation to previous work}\label{sec:std}
The straight linear sum has an interesting connection to what is known as `standard'  reconstruction \cite{Eisenstein:2006nk} (see \cite{rec1loop} for a perturbative analysis and \cite{cp2023} for a summary of related methods).  In this approach, the observed nonlinear, biased field $\delta_b(\bm{k})$ is first smoothed with a filter $G(|\bm{k}|)$, and then used to define a displacement field $\bm{S}_G(\bm{k})\equiv {\rm i}\,(\bm{k}/k^2)\,[\delta_b(\bm{k})/b] \,G(k)$.  The observed particles are shifted by $-\bm{S}_G(\bm{k})$, and the overdensity field which results is called $\delta_{\rm displaced}$.  In addition, a uniform distribution of particles covering the same geometry is also shifted by $-\bm{S}_G(\bm{k})$, and the resulting field is called $\delta_{\rm shifted}$.  The reconstructed field is $\delta_{\rm rec} \equiv \delta_{\rm displaced} - \delta_{\rm shifted}$. This method requires knowledge of $b$ (hence $\sigma_8$), and depends on a smoothing filter $G(k)$ to satisfy the linearized version of the continuity equation.  Typically $G(k)\approx \exp(-k^2R^2/2)$ with $R\approx 15h^{-1}$Mpc or more, mainly calibrated with simulations.  If $\delta_{\rm displaced}\to 0$, then $\delta_{\rm rec} \approx - \delta_{\rm shifted}$.  If, in addition, these shifts are gradients of a potential field (more likely if the smoothing scale $R$ is large), then $\delta_{\rm shifted}\approx \bm{\nabla}\cdot\bm{S}_G$, which is the moral equivalent of $\Theta_{\rm OT}$ (modulo the smoothing window).  

The correspondence is even closer.  The `standard' method assumes that 
 $\delta_m = \delta_b/b$ (on sufficiently large scales) 
whereas OT models the `dust'.  However, for the simplest Wiener filter dust model (equation~\ref{eq:wienerdust}), if $f_b$ is the mass fraction in biased tracers, then the total mass fluctuation field is given by 
\begin{align}
    \delta_m &\equiv f_b\,\delta_b + (1-f_b)\,\delta_d \nonumber\\
    &\approx [f_b + (1-f_b)\,(bb_d/b^2)]\,\delta_b  
    = \delta_b/b,
\end{align}
where the second line assumes that $\delta_d/b_d = \delta_b/b = \delta_m$ on BAO scales (i.e. shot-noise can be ignored), and the final expression follows from the fact that $f_bb + (1-f_b)\,b_d \equiv 1$. Therefore, unless more sophisticated models for the dust are used, the starting assumptions for OT and the standard approaches are very similar.  Of course, OT is {\em guaranteed} to have $\delta_{\rm displaced}=0$ without having to introduce any smoothing window.  And, perhaps more importantly, as Ref.~\cite{OTrsd} notes, OT is able to quantify the systematic error which comes from inaccuracies in the model for the dust.

The OT approach also has a close connection to the work of Ref.\cite{gaussSeidel}.  We discuss this in Appendix~\ref{sec:gs}.  OT differs from this and more standard approaches by explicitly tracking the `dust' component separately.  As a result, the displacements it returns for the biased tracers are physically intuitive:  e.g., for halos, OT returns both how halos have moved, and how their shapes evolved as they formed.  In addition, it returns at least two reconstructed fields:  the halos and the dust.  In the previous section, we showed that this allowed us to provide a reconstructed BAO signal that had higher signal to noise than if we had worked with either field individually.  We exploit another benefit of having more than one field in the next section.

\section{Multiple tracer analyses}\label{sec:multi}
Although we have focused on the virtues of OT reconstruction for BAO-related science, there is considerable interest in using the large scale clustering signal to constrain the statistical properties of the initial fluctuation field.  

\begin{figure}
    \centering
    \includegraphics[width=\linewidth]{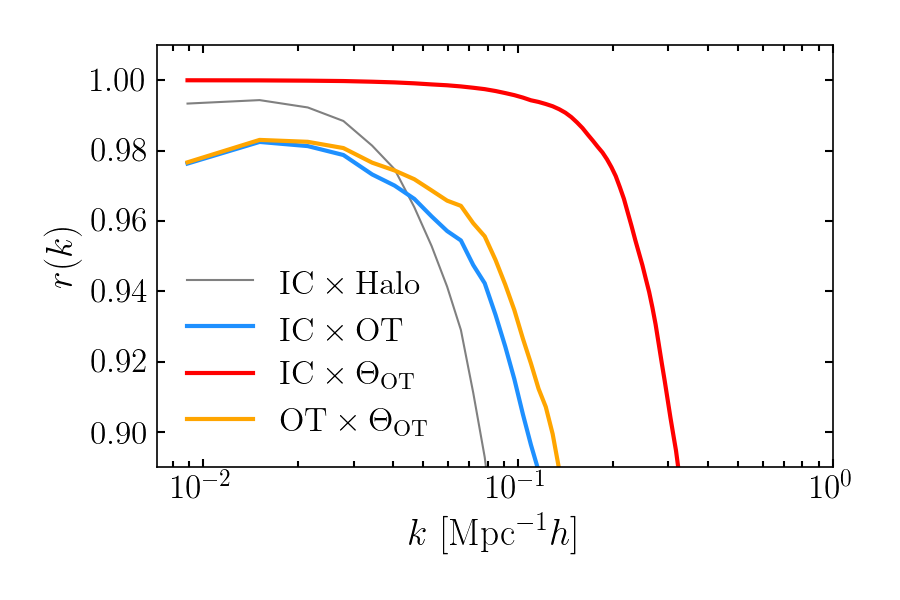}
    \caption{Correlation coefficients between the initial field and the $\Theta_{\rm OT}$ estimate of it, the reconstructed halo field, and the evolved (unreconstructed) halo field, and between the two reconstructed fields. Note that $r(k)$ may not be 1 at large scales due to shot noise.
    \label{fig:rks}}
\end{figure}

For example, suppose we wish to find that linear combination of the reconstructed halo and $\Theta$ fields, $\delta_h$ and $\Theta_{\rm OT}$, for which the difference from the full (initial) matter field $\delta_m$ is as small as possible (in a least squared sense).  Then, we want that $w_h$ and $w_\Theta$ which minimize 
\begin{equation}
 \chi^2(k)\equiv \sum_i \Big(\delta_m(\bm{k}) - w_h\, \delta_h(\bm{k}) - w_\Theta\,\Theta_{\rm OT}(\bm{k})\Big)^2 ,
\end{equation}
where the sum is over all modes with magnitude $|\bm{k}|=k$.  This yields 
\begin{align}
    w_h &= \frac{P_{mh}}{P_{hh}}
    \frac{1 - r_{h\Theta}^2 \,(r_{m\Theta}/r_{mh}r_{h\Theta})}{1 - r_{h\Theta}^2} \nonumber\\
    w_\Theta &= \frac{P_{m\Theta}}{P_{\Theta\Theta}}
    \frac{1 - r_{h\Theta}^2\,(r_{mh}/r_{m\Theta}r_{\Theta h})}{1 - r_{h\Theta}^2}
    \label{eq:wopt}
\end{align}
with variance 
\begin{align}
    \chi^2_{\rm min}  
    &= P_{mm}\left(1 - r_{mh}^2 - \frac{(r_{m\Theta} - r_{mh}r_{h\Theta})^2}{1 - r_{h\Theta}^2}\right)\nonumber\\
    &= P_{mm}\left(1 - r_{m\Theta}^2 - \frac{(r_{mh} - r_{m\Theta}r_{h\Theta})^2}{1 - r_{h\Theta}^2}\right).
\end{align}
If we only used one field, $\delta_h$, then the variance would be $P_{mm}(1 - r_{mh}^2)$, so the final term shows how much the variance is reduced by adding the extra field.  There is no reduction only if the correlation between $\delta_m$ and $\Theta$ is entirely due to the correlations of $\delta_m$ and $\delta_h$ with $\Theta$:  i.e., only if $r_{m\Theta} = r_{mh}r_{h\Theta}$.  Figure~\ref{fig:rks} shows the various cross-correlation coefficients:  $r_{m\Theta} \ne r_{mh}r_{h\Theta}$ in general.  

\begin{figure}
    \centering
    \includegraphics[width=\linewidth]{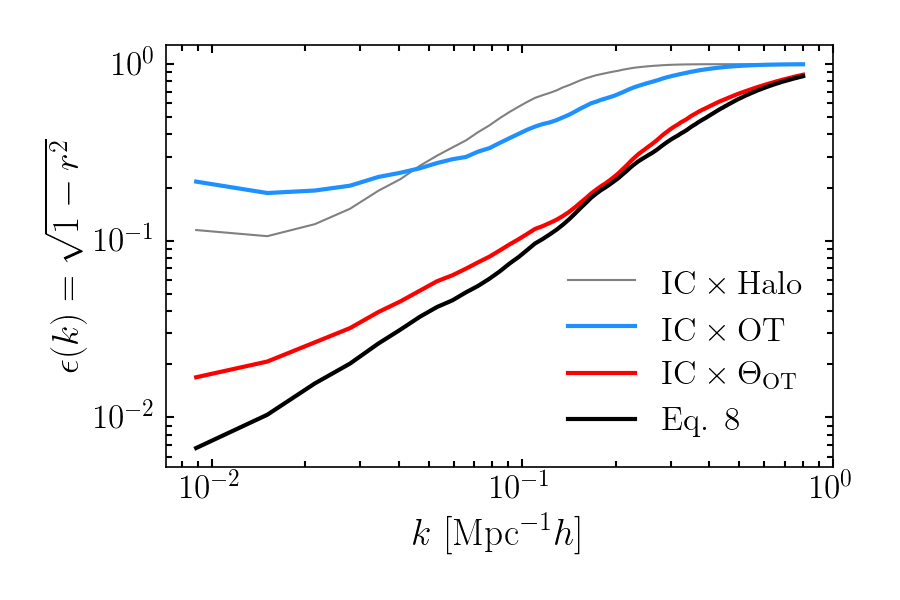}
\caption{Stochasticity between the initial fluctuation field and the evolved halo field, the OT reconstructed halos, the divergence of the OT displacements of the full field, and the optimal linear combination of the two reconstructed fields.
    \label{fig:Ek}}
\end{figure}

It is conventional \cite[e.g. Ref.][]{cai2011} to use $\sqrt{\chi^2_{\rm min}/P_{mm}}$ as a measure of the stochasticity between the optimally weighted fields, $w_h\delta_h + w_\Theta\Theta$, with $w_h$ and $w_\Theta$ given by equation~(\ref{eq:wopt}), and the one they are trying to represent, $\delta_m$.  Figure~\ref{fig:Ek} shows the stochasticity as a function of $k$. For this (mass-weighted) halo example, the reconstructed halos are closer to the ICs than the evolved halos -- reconstruction was useful -- but the reconstructed $\Theta_{\rm OT}$ field is much closer, so that adding $\Theta_{\rm OT}$ to the halos results in significant gains (i.e. reduction in stochasticity), but not vice versa.

Furthermore, because OT reconstruction does not assume that the initial field was Gaussian, it provides a useful framework for studying departures from Gaussianity. Indeed, Ref.~\cite{2006MNRAS.365..939M} explored the use of the one-point distribution of divergences in the full OT reconstructed field (i.e. without considering the added complication of starting from a biased tracer field) as a tool for quantifying primordial non-Gaussianity. However, other constraints on primordial non-Gaussianity are now sufficiently tight that one-point statistics are not expected to provide competitive constraints; two-point or higher-order statistics are expected to be more informative.

Models where the primordial non-Gaussianity is `local' are a case in point.  These have $\phi(\bm{x}) = \phi_{\rm G}(\bm{x}) + f_{\rm NL}\, \phi_{\rm G}^2(\bm{x})$, where $\phi_{\rm G}$ is an isotropic Gaussian random field.  
 On large scales in these models, biased tracers satisfy 
 $\delta_b(\bm{k}) = b_\delta\delta(\bm{k}) + f_{\rm NL}\, b_\phi \phi(\bm{k})$.  
Since $\phi(\bm{k})\propto \delta(\bm{k})/k^2$, the clustering of biased tracers has a characteristic scale-dependence \cite{PNGdalal}.  However, the scales on which this is dramatic are so large that cosmic variance on the measured power spectrum is a concern.  On the other hand, the cosmological information here -- in this case the amplitude of $f_{\rm NL}$ -- is carried, not by the power spectrum itself, but by the ratio of the biased and unbiased fields, or by the ratio of two differently biased fields.  

If both fields suffer the same cosmic variance then, to first approximation, cosmic variance cancels when one studies their ratio \cite{multitracer}.  To be useful in practice, the tracer fields must occupy the same survey volume {\em and} have different bias factors.  For two such tracers, $A$ and $B$, the error on $f_{\rm NL}$ is proportional to $|b_\delta^A b_\phi^B - b_\delta^B b_\phi^A|^{-1}$ \cite{BKfNL}.  If $b_\phi\propto b_\delta-1$ then this becomes $(b_\delta^B - b_\delta^A)^{-2}$.  We will work with this approximation in what follows, but it is not crucial to the spirit of our analysis.  

The main point we wish to make is that OT reconstruction naturally provides us with at least three tracer fields, in addition to the observed, unreconstructed one, which can be used to address such problems:  the mass-weighted and unweighted halo reconstructed fields, and $\Theta$ (not to mention linear combinations of these fields).  Roughly, such studies require 
 $\bar{n} (b_\delta^B - b_\delta^A)^2 P_{mm} \gg 1$.  
If $b_\delta^B$ and $b_\delta^A$ refer to halos before and after reconstruction, then $b_\delta^A = b_\delta^B-1$, so $b_\delta^B-b_\delta^A=1$, and we require $\bar{n} P_{mm} \gg 1$.  Typically, surveys have $\bar{n}\, (b_\delta^B)^2 P_{mm} \sim 3$, so we would like a large enough number density of tracers that $\bar{n}\, P_{mm} \gg (3/(b_\delta^B)^2)$.  (E.g., in Figure~\ref{fig:PkGz}, the effective number density is $\sim 10^{-4}$ so the (un)reconstructed halos have $\bar{n}b^2P\sim (10)3$.)  Of course, sample $B$ could be mass-weighted halos at their evolved positions, and sample $A$ could be the unweighted halos at their reconstructed positions:  this would increase $b_\delta^B - b_\delta^A$, and so improve the constraint on $f_{\rm NL}$.  

In this context, it is worth mentioning that there exists a well-developed literature on how to best weight biased tracers such as halos so as to reproduce the full matter distribution with as little stochasticity as possible \cite{fNLoptimal, cai2011}.  For halos, the optimal weight (in a linear least-squares sense), is the sum of two terms:  the halo mass and a Wiener-filter like term which involves the halo-dust cross-correlation function.  Since OT works most easily with mass-weighted halos, it would be natural to apply these optimal weighting schemes to the evolved {\em and} reconstructed halo fields, before combining them in a multi-tracer analysis.  We leave this to future work.  

\section{Discussion and conclusions}\label{sec:discussion}
Optimal Transport reconstruction returns the set of displacements which map a clustered point distribution to a uniform field.  If the displacements are gradients of a potential, they are exactly reconstructed.  Although the reconstructed point distribution is uniform, the field defined by the divergence of the displacements, $\Theta(\bm{q})$ is not.  We showed that this field is quite well-correlated with the initial density fluctuation field, evolved using linear theory (purple curve in Figure~\ref{fig:rk}).  I.e., $\Theta(\bm{q})$ represents the OT reconstruction of the linearly evolved field starting from the non-linearly evolved field, even though OT makes no assumption about the amplitude of linear theory growth.  We argued that because OT reconstructs displacements rather than densities or velocities, 
there is an upper limit to the fidelity with which $\Theta(\bm{q})$ can trace the linear-theory overdensity field (red or green curves in Figure~\ref{fig:rk}).  Nevertheless,  the pair correlation function of this OT reconstructed field, 
 $\xi(|\bm{q}_i-\bm{q}_j|)\equiv \langle\Theta(\bm{q_i})\Theta(\bm{q_j})\rangle$,
is extremely similar to that of the pair correlation function in linear theory $\xi_{\rm Lin}(|\bm{q}|)$, both in amplitude and shape (Figure~\ref{fig:xir}), on the large scales that are relevant to BAO physics.  This is also true of its power-spectrum:  $P_{\Theta\Theta}(|\bm{k}|)\approx P_{\rm Lin}(|\bm{k}|)$ (Figure~\ref{fig:Pk}).  

In practice, only observations of the evolved positions of a subset of the field will be available, and these will suffer from redshift-space distortions.  Previous work has shown how to estimate the distribution of the remaining mass, the `dust', from the observed subset, before the OT reconstruction can be performed.  In principle, inaccuracies in this estimate of the dust can impact the fidelity of the OT reconstructions of the biased tracers.  In practice, even the simplest Wiener filter model of the dust allows rather good reconstruction of the displacements of the the biased tracers.  This remains true for the full field (biased tracers plus dust) as well: although the fidelity of the reconstruction is not as good as in the ideal case, it is still rather good on BAO scales (compare orange and blue curves in Figure~\ref{fig:rkWGz}).  In particular, the reconstructed power spectra and correlation functions are in good agreement with the linear theory shapes and amplitudes, both for the biased subset (yellow and green curves in Figures~\ref{fig:PkGz} and~\ref{fig:xiGz} agree) and for the full field (purple and pink curves agree), on BAO scales.

Previous OT work has made the point that the reconstructed BAO feature in the clustering signal of the biased subset can be used to estimate the cosmological distance scale \cite{PRLhalos}.  Our work now allows the full reconstructed field to be used as well.  Since the two fields are correlated with one another, this raises the question of how best to combine the two.  The amplitude of the correlation function in either of the two fields is smaller than in the original nonlinear field (compare orange and yellow curves with blue in Figures~\ref{fig:PkGz} and~\ref{fig:xiGz}).  However, the signal in the field that is a simple sum of the two fields has the same amplitude as the original measurement, but with an enhanced BAO feature (red curves in Figures~\ref{fig:PkGz} and~\ref{fig:xiGz}). We discussed why summing the two fields is physically reasonable, and why peaks theory provides a useful framework for understanding why this enhances the BAO feature:  peaks theory explains both this enhancement and the fact that the OT displacements for the halo field are slightly smaller than for the total field. Finally, we discussed how this summed field, and our treatment of the `dust', compare to other reconstruction methods in the literature (Section~\ref{sec:std}).  

It is encouraging that the OT methodology can be used to estimate the linear theory correlations.  Nevertheless, there is room for improvement. We have assumed that divergence of OT displacement is a good estimator for linear theory density, but it may be possible to do better.  In addition, our current Wiener filter model for the dust degrades the fidelity of the reconstruction (orange curve lies below the blue in Figure~\ref{fig:rkWGz}).  This `linear' dust model can be improved by including the fact that the dust-tracer correlation depends on the mass of the tracers and their environment \cite{cai2011, cai2023}.  Alternatively, recent machine learning-based methods may provide better `nonlinear' dust assignment models \cite{AEdust, CNNdust}.  Incorporating these improvements into our approach is the subject of work in progress.

We believe there is good reason for doing so.  Some cosmological analyses benefit from having multiple tracer fields over the same survey volume (Section~\ref{sec:multi}).  OT reconstruction naturally provides us with at least four tracer fields, in addition to the observed, unreconstructed one:  the unweighted, mass- and optimally-weighted reconstructed halo fields, and $\Theta$ (not to mention combinations of these fields).  
For such studies, the inclusion of higher fidelity dust models and optimally weighting the reconstructed fields should reap rich dividends.

\acknowledgements
We are grateful to the organizers and participants of the workshop on Optimal Transport held at Les Houches in March 2023 where some aspects of this work was discussed.  BL, RM and FN thank the Rudolf Peierls Centre for Theoretical Physics at Oxford University for hospitality, and RKS is grateful to the ICTP for its hospitality during the summer of 2023. FN gratefully acknowledges support from the Yale Center for Astronomy and Astrophysics Prize Postdoctoral Fellowship. NP is supported in part by DOE DE-SC0017660.

\appendix
\section{Technical details about estimating divergences}

In the initial conditions of the simulations, we have $\bm{v}(\bm{r})$ on a grid in configuration space.  We do not estimate the divergence in configuration space. Instead, we perform a Fast Fourier Transform (FFT) to obtain $\bm{v}(\bm{k})$ on a grid in Fourier space.  We then evaluate ${\rm i}\bm{k}\cdot \bm{v}(\bm{k})$ at each Fourier space grid point and divide by $(afH)_{init}$ to get $\Theta_{\rm Zel}(\bm{k})$. Finally, if needed, we inverse-FFT back to real space. The power spectrum of the divergence is determined by summing the squared magnitudes of the components of $\Theta_{\rm Zel}(\bm{k})$ over all available $\bm{k}$ modes.

In practice, for OT-reconstructions, we estimate the divergence by interpolating the displacements $\bm{S}_{\rm OT}(\bm{r})$ of the reconstructed Laguerre cell barycenters onto a real-space grid. We then perform an FFT to obtain $\bm{S}_{\rm OT}(\bm{k})$ on a grid in Fourier space, evaluate ${\rm i}\bm{k}\cdot \bm{S}_{\rm OT}(\bm{k})$ at each Fourier space grid point, and finally, compute the power spectrum.  However, in sparse tracer fields, there may be many empty voxels.  Assigning zero velocity to such voxels is incorrect. To address this in our OT reconstruction measurements, we compute the divergence of both halo and dust displacements, since this combined field is not sparse. We also perform different convergence tests, such as varying grid sizes and mass assignment schemes for the displacement field, to effectively mitigate this issue. 

For our analyses, we chose a three-dimensional grid with 512 points along each dimension, resulting in a total $512^3$ grid points or voxels. We also used the Triangular Shaped Cloud (TSC) mass-assignment scheme for assigning mass (or displacement values) to the grid points. The TSC scheme is a higher-order mass assignment method that distributes the mass of particles over a local neighborhood of grid points, using a weighted average where the weights decrease with distance from the particle's actual position. Compared to simpler schemes such as Nearest Grid Point (NGP) or Cloud-in-Cell (CIC), TSC offers improved accuracy in representing the density and displacement fields by considering a larger neighborhood around each point. This smoother representation of the field is also advantageous in Fourier space analysis and in reducing numerical artifacts.

\begin{figure}
    \centering
    \includegraphics[width=\linewidth]{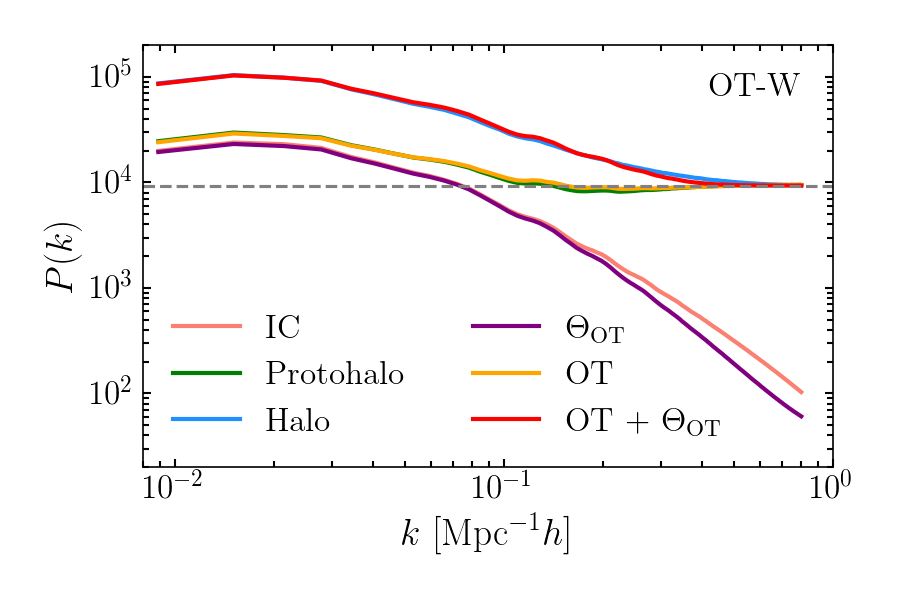}
    \caption{Same as Figure~\ref{fig:PkGz} but when the OT algorithm starts with the (point-mass) halos in real space and the correct distribution of dust.  
    \label{fig:PkW}}
\end{figure}

\begin{figure}
    \centering
    \includegraphics[width=\linewidth]{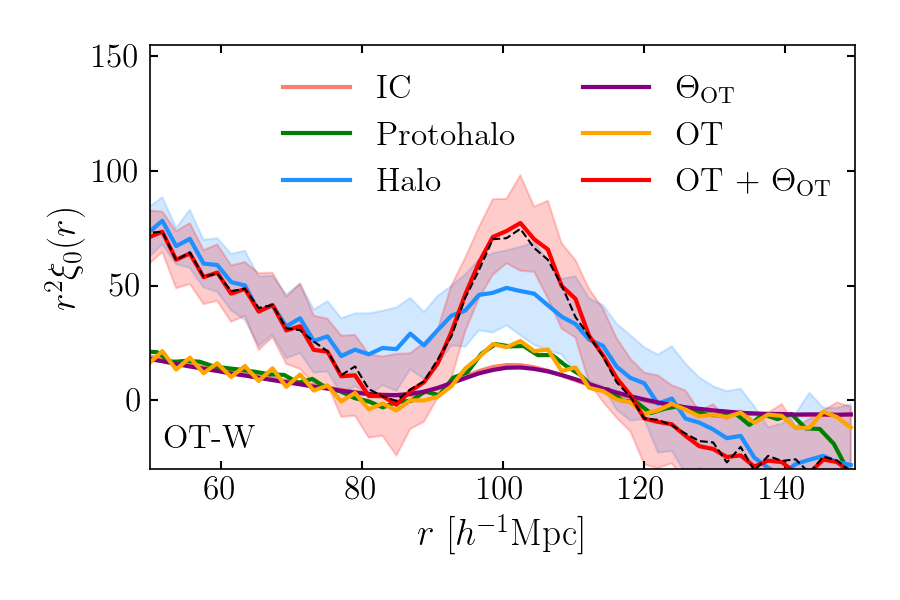}
    \caption{Correlation functions corresponding to the power spectra shown in Figure~\ref{fig:PkW}. The dashed black curve shows the correlation function of the sum of the actual protohalo and the initial fluctuation fields.
    \label{fig:xiW}}
\end{figure}

\section{Effect of point-mass halos only}\label{sec:otw-only}
The main text showed results which reconstruct the initial protohalo and matter fields starting from biased tracers in redshift space, with a simple Wiener filter model for the dust.  This model leads to the lowest of the three $r(k)$ curves shown in Figure~\ref{fig:rkWGz}.  The middle curve there starts from the same halos, but in configuration space, and uses the true dust.  This cannot be done in real data, but, for completeness, Figures~\ref{fig:PkW} and~\ref{fig:xiW} show the associated $P(k)$ and $\xi(r)$.  In contrast to Figures~\ref{fig:PkGz} and~\ref{fig:xiGz}, the reconstructed (red) and input (blue) curves have the same amplitude (because the blue curve is not redshift-space distorted).

\section{Connection to reconstruction using multigrid Gauss-Seidel relaxation}\label{sec:gs} 

Ref.\cite{gaussSeidel} describes a multigrid Gauss-Seidel relaxation method for reconstructing the initial field from the evolved one. We discuss how it compares to our method here. Related to this discussion, and as mentioned in the text, the displacement can be equivalently thought of as a function of $\bm{q}$ or $\bm{x}$. Here, we also show explicitly the difference between taking the derivative with respect to $\bm{x}$ and $\bm{q}$.

\subsection{Displacement potential in the initial field}
We start with the gradient of equation~(\ref{eq:xqSq}), because it is related to the nonlinear density:
\begin{equation}
  1+\delta(\bm{x}) = \frac{1}{|d\bm{x}/d\bm{q}|} .
  \label{eq:invdxdq}
\end{equation} 
If the displacements $\bm{S}$ are gradients of a potential $\psi(\bm{q})$, then the nonlinear density is the (inverse of the) Jacobian determinant of the `deformation tensor' of second derivatives of the potential.  In this case, we can write the determinant as 
\begin{equation}
  \frac{d\bm{x}}{d\bm{q}} = {\rm det}|\delta_{ij} - \psi_{ij}| = 1 - I_1 + I_2 - I_3 ,
  \label{eq:dxdqIs}
\end{equation}
where 
\begin{align}
 I_1 &= {\rm Tr}(\psi_{ii}) = \sum_i \psi_{ii} = -\bm{\nabla}_{\bm{q}}\cdot \bm{S}, \nonumber\\
 I_2 &= \sum_{j>i} \psi_{ii}\psi_{jj} - \psi_{ij}\psi_{ij}, \\
 I_3 &= {\rm det}(\psi_{ij}).\nonumber
\end{align}
Thus, 
\begin{equation}
  1+\delta(\bm{x}) = \frac{1}{|d\bm{x}/d\bm{q}|} 
  = 1 + I_1 + \frac{2}{3}\, I_1^2 + \frac{I_1^2 - 3I_2}{3} + \ldots ;
 \label{eq:dNLIs}
\end{equation} 
although the nonlinear density is given by the determinant of the deformation tensor, the leading order term, $I_1$, is the trace of the deformation tensor.  It is useful to think of this trace as the sum of the three eigenvalues of the tensor, because
\begin{equation}
    \bm{\nabla}_{\bm{q}}\cdot \bm{S} 
    = \bm{\nabla}_{\bm{q}}\cdot \bm{x} - 3
    = \sum_{i=1}^3 (1-\lambda_i) - 3 = -\sum_{i=1}^3 \lambda_i .
  \label{eq:divSq}
\end{equation}
I.e., the trace equals the divergence (at $\bm{q})$ of the displacements, which is why this divergence played a leading role in the main text.  Note that the divergence evaluated at $\bm{x}$ (rather than $\bm{q}$) of the same displacements is 
\begin{equation}
    \bm{\nabla}_{\bm{x}}\cdot \bm{S} 
    = 3 - \sum_{i=1}^3 (1-\lambda_i)^{-1} = -\sum_{i=1}^3 \frac{\lambda_i}{1-\lambda_i}.
\label{eq:divSx}
\end{equation}
This `nonlinear' divergence diverges as $\lambda_i\to 1$.

The combinations 
\begin{equation}
    \delta_{\rm L}\equiv I_1,\quad 
    q_{\rm L}^2 \equiv I_1^2 - 3I_2,\quad
    u_{\rm L}^3 \equiv \frac{2I_1^3}{9} - I_1I_2 + 3I_3
 \label{eq:dqu}
\end{equation}
are special.  If we define 
\begin{equation}
  \overline{\lambda}_i \equiv \lambda_i - \frac{\sum_i\lambda_i}{3} 
\end{equation}
then 
\begin{equation}
 \delta_{\rm L} = \sum_i\lambda_i,\quad
 q_{\rm L}^2 = \frac{\sum_i\overline{\lambda}_i^2}{2/3} \quad{\rm and}\quad
 u_{\rm L}^3 = \sum_i \overline{\lambda}_i^3 .
\end{equation} 
The analysis above shows that $\delta_{\rm L}$ equals the divergence (at the initial position $\bm{q}$) of displacements 
 $\delta_{\rm L} = -\bm{\nabla}_{\bm{q}}\cdot \bm{S}$.  The quantity 
$q_{\rm L}^2$ is the `traceless shear' \cite[e.g.][]{nonlocalBias}.  
In a Gaussian random field, $\delta_{\rm L}$ is a Gaussian variable, $q_{\rm L}^2$ is distributed as $\chi^2_5$ and is independent of $\delta_{\rm L}$, and $\mu_{\rm L} \equiv (9 u_{\rm L}^3/2)/q_{\rm L}^3$ is distributed like the cosine of an angle:  it is uniform over $-1\le\mu_{\rm L}\le 1$.  

In terms of these variables, equation~(\ref{eq:dxdqIs}) reads
\begin{equation}
  \frac{d\bm{x}}{d\bm{q}} 
   = (1 - \delta_{\rm L}/3)^3\,\Big(1 - Q_{\rm L}^2/3 - U_{\rm L}^3/3 \Big) ,
\end{equation}
where 
 $Q_{\rm L}\equiv q_{\rm L}/(1 - \delta_{\rm L}/3)$ 
and 
 $U_{\rm L}\equiv u_{\rm L}/(1 - \delta_{\rm L}/3)$.
This form will be useful below, in part because $q_{\rm L}$ and $u_{\rm L}$ both vanish in spherical symmetry (when all the $\lambda_i$ are equal then all $\overline{\lambda}_i=0$), so the first term isolates the spherical contribution, and the second term encodes the corrections from asphericity.

\subsection{Displacement potential in the displaced field}
Ref.\cite{gaussSeidel} assume that 
\begin{equation}
    1+\delta_{\rm NL}(\bm{x}) = \frac{d\bm{q}}{d\bm{x}}
    = {\rm det}|\delta_{ij} + \zeta_{ij}|
  \label{eq:dqdx}
\end{equation}
where $\bm{q} = \bm{x}-\bm{S}$.  The difference in notation compared to equation~(\ref{eq:invdxdq}) is intended to highlight the fact that here we will work with derivatives with respect to $\bm{x}$, whereas until now all derivatives were with respect to $\bm{q}$.  

The second equality follows from assuming that the displacements $\bm{S}$ are gradients of a potential $\zeta(\bm{x})$, so the determinant on the right hand side involves second derivatives (wrt the evolved position $\bm{x}$) of $\zeta(\bm{x})$.  However, if the flow is potential, then the displacement $\bm{S}=\bm{x}-\bm{q}$ can also be thought of as the gradient with respect to the initial position $\bm{q}$ of a (different) potential $\psi(\bm{q})$:  In particular, 
 $\nabla_{\bm q}\psi = \bm{S} =\nabla_{\bm x}\zeta$, 
and the potentials $\zeta(\bm{x})$ and $\psi(\bm{q})$ are Legendre transforms of one another \cite{EUR}. The main point we make below is that Ref.\cite{gaussSeidel} aims to estimate $\zeta(\bm{x})$ whereas we estimate $\psi(\bm{q})$ (compare equations~\ref{eq:dqdx} and~\ref{eq:invdxdq}).  

To see how the methods are related, let $1 + \Lambda_i$ denote the eigenvalues of the deformation tensor $\delta_{ij} + \zeta_{ij}$.  In terms of these, equation~(8) of Ref.~\cite{gaussSeidel} reads 
\begin{equation}
    1+\delta_{\rm NL} = (1 + \nabla^2\zeta/3)^3\,\Big(1 - Q_{\rm NL}^2/3 + U_{\rm NL}^3/3 \Big) ,
\end{equation}
where 
 $Q_{\rm NL}$ and $U_{\rm NL}$ are the same combinations of the $\Lambda_i$ that 
$Q_{\rm L}$ and $U_{\rm L}$ are of the $\lambda_i$. A little algebra shows that, for the right hand sides of equations~(\ref{eq:dqdx}) and~(\ref{eq:invdxdq}) to equal one another, 
\begin{equation}
    \Lambda_i(\bm{x}) = \frac{\lambda_i(\bm{q})}{1-\lambda_i(\bm{q})}.
  \label{eq:Lili}
\end{equation}
This is just as expected from equation~(\ref{eq:divSx}), and is most easily shown by using equation~(\ref{eq:Lili}) in their equation~(3).

Since the quantity of interest for reconstruction is $\delta_{\rm L}$, Section II.B of Ref.~\cite{gaussSeidel} describes the additional steps that must be taken to estimate $\delta_{\rm L}(\bm{q})$ once $\zeta(\bm{x})$ is estimated.  However, equation~(\ref{eq:Lili}) shows that this is not really necessary, since 
\begin{equation}
  \delta_{\rm L}(\bm{q})\equiv \sum_i\lambda_i(\bm{q}) = \sum_i \frac{\Lambda_i(\bm{x})}{1+\Lambda_i(\bm{x})},
\end{equation}
where it is understood that $\bm{x}$ is displaced from $\bm{q}$ by the gradient of a potential {\it i.e.} their analysis does not exploit the fact that $\zeta(\bm{x})$ and $\psi(\bm{q})$ are related by a Legendre transform.

The analysis of Ref. \cite{gaussSeidel}, like that of Ref.\cite{PRLdm}, assumes knowledge of the full evolved density field $\rho(\bm{x})$.  To treat the (more realistic) case in which positions of only a biased subset are known, Ref.~\cite{gaussSeidelHalos} argue that one must assume a model for the bias (e.g. $\delta_b = b_1\delta + b_2 \delta^2/2 + \ldots $), and use this to express $\delta$ in terms of $\delta_b$.  In effect, this assigns weights to the halo field.  This differs from our approach, in which we assume a simple model for the `dust' (essentially, $\delta_b = b_1\delta$; while it is potentially interesting to extend the dust model to incorporate high-order bias models, this would come with the cost of introducing more free parameters).  Moreover, as Ref.~\cite{PRLhalos} notes, because we use `dust' rather than a modified weight at the location of each halo, our approach accurately recovers the displacement vectors of halos, as well as the evolution of their shapes, extremely well.  This matters because the protohalo centers are {\em not} uniformly distributed in the initial conditions, so an approach which weights the halo field will have trouble returning a displacement field which is physically intuitive.

An additional benefit of our methodology is that, in contrast to Gauss-Seidel relaxation, the OT algorithm is guaranteed to converge to the unique solution \cite{KMT2019}. In Ref. \cite{gaussSeidel}  BAO reconstruction problem was studied also because of its mathematical similarities to a particular class of modified gravity theories :  the field equation in the `quartic Galileon model' is also a nonlinear elliptical partial differential equation.  Having demonstrated the correspondence between the semi-discrete OT algorithm and the Gauss-Seidel relaxation one, our work suggests that it would be interesting to see if semi-discrete OT is useful in the context of quartic Galileons, especially because Gauss-Seidel has linear speed of convergence whereas for OT it is nearly quadratic.

\bibliography{OTdiv.bib}

\end{document}